\documentclass[a4paper,
               ]{jacow}
\usepackage{enumitem}
\usepackage{varwidth}
\usepackage{pdfpages,multirow,ragged2e} %
\makeatletter%
	\ifboolexpr{bool{xetex}}
	 {\renewcommand{\Gin@extensions}{.pdf,%
	                    .png,.jpg,.bmp,.pict,.tif,.psd,.mac,.sga,.tga,.gif,%
	                    .eps,.ps,%
	                    }}{}
\makeatother

\ifboolexpr{bool{xetex} or bool{luatex}} 
 {}                                      
 {\usepackage[utf8]{inputenc}}           

\usepackage[USenglish]{babel}

\ifboolexpr{bool{jacowbiblatex}}%
 {%
  \addbibresource{jacow-test.bib}
  \addbibresource{biblatex-examples.bib}
 }{}
\begin{document}

\title{Temperature, rf field, and frequency dependence performance evaluation of superconducting niobium half-wave coaxial cavity \thanks{ Work supported by the U.S. Department of Energy, Office of Science, Office of Nuclear Physics under contract DE-AC05-06OR23177.}}

\author{N. K. Raut\textsuperscript{1}\thanks{raut@jlab.org}, B. D. Khanal\textsuperscript{2}, J. K. Tiskumara\textsuperscript{2}, S. De Silva\textsuperscript{2}, \\
P. Dhakal\textsuperscript{1}, G. Ciovati\textsuperscript{1,2}, and J. R. Delayen\textsuperscript{1,2} \\
\textsuperscript{1}Jefferson Lab, Newport News, VA 23606, USA \\
\textsuperscript{2}Center for Accelerator Science and Department of Physics, Old Dominion University, \\
Norfolk, VA, 23529, USA}
	
\maketitle

\begin{abstract}
  Recent advancement in superconducting radio frequency cavity processing techniques, with diffusion of impurities within the RF penetration depth, resulted in high quality factor with increase in quality factor with increasing accelerating gradient. The increase in quality factor is the result of a decrease in the surface resistance as a result of nonmagnetic impurities doping and change in electronic density of states. The fundamental understanding of the dependence of surface resistance on frequency and surface preparation is still an active area of research. Here, we present the result of RF measurements of the TEM modes in a coaxial half-wave niobium cavity resonating at frequencies between 0.3 – 1.3 GHz. The temperature dependence of the surface resistance was measured between 4.2 K and 1.6 K. The field dependence of the surface resistance was measured at 2.0 K. The baseline measurements were made after standard surface preparation by buffered chemical polishing. 
\end{abstract}

\section{Introduction}
Superconducting radio-frequency (SRF) cavities are the building blocks of modern particle accelerators that can store and transfer electromagnetic energy with very little dissipation \cite{Padamsee}. High-quality factor in SRF cavities is not only limited to particle accelerators but also emerging as an application to quantum computing and quantum information science \cite{krasnok, Raut, dhakalJNPS} . 

The performance of SRF cavities is measured in terms of the quality factor ($Q_0$) which is inversely proportional to the surface resistance ($R_s$) as a function of accelerating gradient ($E_{acc}$). Surface engineering with nonmagnetic impurity doping (Ti, N, O) has been shown to reduce surface resistance and improve the quality factor \cite{Dhakal,Lechner, Ciovati}.  Nevertheless, the fundamental understanding of the dependence of surface resistance on frequency, surface preparations, and RF field is still an active area of research \cite{Park}. In this contribution, we have measured the frequency, temperature, and RF field dependence of surface resistance using a half-wave coaxial cavity. The benefit of using the same cavity for different modes will eliminate the variability that comes when using different cavities, although the same treatment is applied.


\section{Cavity Design and Surface Preparation}
\begin{figure*}[!htb]
   \centering
   \includegraphics*[width=\textwidth]{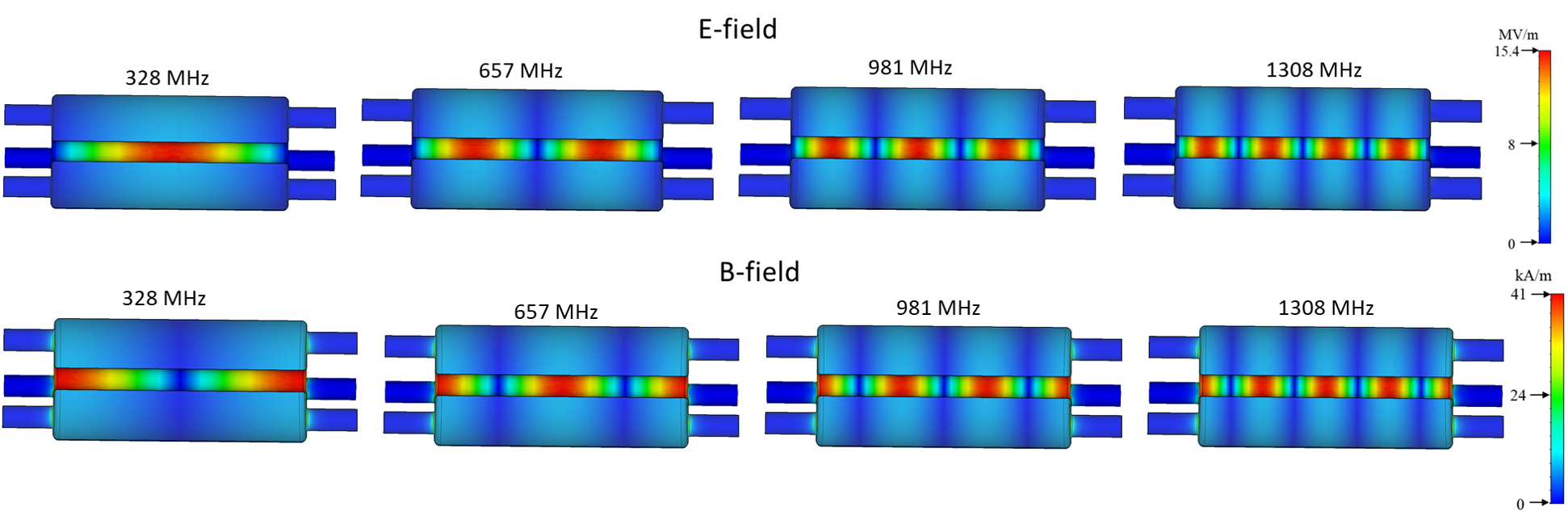}
   \caption{Electric and magnetic field distribution of the first four TEM modes of the half-wave coaxial cavity. The E-field oscillates between the inner and outer conductors of the cavity and the B-field circulates around the center conductor. }
   \label{fig:Fig1}
\end{figure*} 
The quality factor is defined as the ratio of energy stored (U) to the energy dissipation ($P_d$) per rf cycle as:
\begin{equation}
 Q_0 =\frac{\omega U}{P_d} = \frac{2\pi f \mu_0\iiint_V |H|^2 dV}{R_s\iint_S |H|^2 dS}  = \frac{G}{R_s}
\end{equation}
\label{eq1}
where, 
\begin{equation}\label{eq:2}
  G=\frac{2\pi f_0 \mu_0\iiint_V |H|^2 dV}{\iint_S |H|^2 dS}
\end{equation}
is geometric factor of the cavity and depends on the shape.
Here, $f_0$, $\mu_0$, and |H| represent the frequency, vacuum permeability, and peak RF magnetic field. $R_s$ represents the surface resistance of the cavity which is expressed as the sum of temperature independent $R_i$ and temperature dependent $R_{BCS}$. The $R_i$ arise due to several intrinsic and extrinsic factors and $R_{BCS}$ is the result of RF dissipation by unpaired quasi-particles in superconductors and is explained by BCS theory of superconductivity \cite{MB}. A simplified expression for $R_{BCS}$ valid in the dirty limit and $T << T_c$ is \cite{Turneaure}: 
\begin{equation}\label{eq:3}
  R_{BCS} (T) \approx \frac{A}{T}f^2e^{(-\frac{\Delta}{k_BT_c}\frac{T_c}{T})}
\end{equation}
where $A$ depends on the material properties like penetration depth, coherence length, Fermi velocity, etc. $\Delta$ is the energy gap at 0 K, $k_B$ is Boltzmann constant, and $T_c$ is the transition temperature. 

Accelerating elliptical cavities are excited in $TM_{010}$ mode where the electric field is concentrated along the cavity axis with peak magnetic field on the surface of the cavity. The fundamental and higher-order transverse electromagnetic wave (TEM) modes of the half-wave coaxial cavity have a strong field on the center conductor that decays exponentially towards the outer conductor \cite{parksrf15, parksrf17, parksrf19}. Such identical field distribution of the TEM modes on the center conductor makes it an ideal platform to study the frequency dependence of the surface resistance at different temperature, RF field and for different surface treatments \cite{Gurevich}. 

The length of the half-wave coaxial cavity used in this experiment is $l=457.55$ mm, which determines the fundamental mode frequency ($f$) by the relation $f=\frac{c}{2l}$. The TEM modes are sufficiently 
separated by other TM and TE modes. Our modes of interest are fundamental and three higher harmonics of the cavity. Figure \ref{fig:Fig1} shows the E and B profile of the first four TEM modes. The CST Studio Suite simulation was done and used to calculate the RF parameters as outlined in Table \ref{table1}. 

\begin{table}[!hbt]
   \centering
   \caption{RF parameters of four TEM modes of the half-wave coaxial cavity. }
   \begin{tabular}{lcc}
       \toprule
\textbf{Modes} & \textbf{f (MHz)}  & \textbf{G ($\Omega$)}\\
       \midrule
TEM1    & {328}   & 61.8    \\ 
TEM2    & {657}   & 123.0   \\ 
TEM3    & {985}   & 186.0   \\ 
TEM4     & {1313}   & 247.0  \\ 
       \bottomrule
   \end{tabular}
   \label{table1}
\end{table}

The cavity was fabricated using the deep-drawing of the parts and electron beam welding at Jefferson Lab. After the fabrication, the cavity was subjected to $\sim$ 150 $\mu$m buffered chemical polishing (BCP), followed by hydrogen degassing at 800 $^\circ$C for 3 hrs in an ultra-high vacuum furnace. After the furnace treatment, the cavity was again subjected to 30 $\mu$m BCP, high-pressure rinse, and clean assembly of the cavity with input and pick-up probes coupled to the magnetic field. The cavity is evacuated and cooldown in Dewar with liquid helium with residual magnetic field in Dewar < 5 mG. The baseline RF measurements consist of the $Q_0$ vs. $T$ measurements at constant RF peak field $B_p \sim$ 10 mT and $Q_0(B_p)$ at 2.0 K.

\section{Experimental Results}
\subsection{RF Results}
\begin{figure}[!htb]
   \centering
   \includegraphics*[width=1\columnwidth]{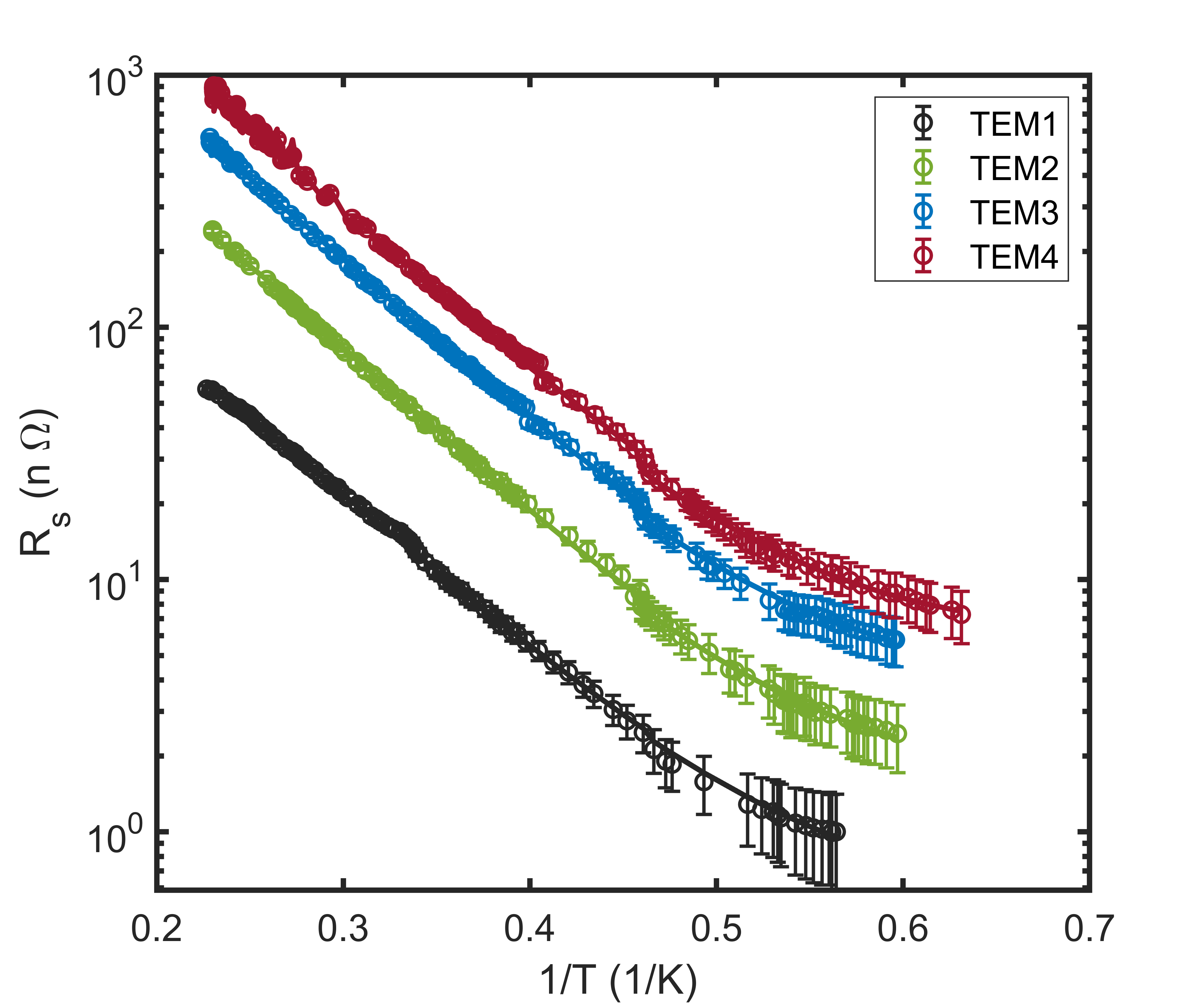}
   \caption{The total surface resistance as a function of temperature for first 4 TEM modes. The lines are the least square fits to extract $R_i$ and $R_{BCS}$.}
   \label{Fig2}
\end{figure}

Figure \ref{Fig2} shows the total resistance ($R_s$) as a function of temperature (1/T) calculated from the $Q_0(T)$ measurement at $B_p \sim$ 10 mT. Solid line represents the least square fit for $R_s = R_i + R_{BCS}$, where $R_{BCS}$ is given by Eq. 3. For all temperature, $R_s$(TEM1)$<$ $R_s$(TEM2)$<$  $R_s$(TEM3)$<$  $R_s$(TEM4).

\subsection{Field Dependence Test at 2 K}
\begin{figure}[!htb]
   \centering
   \includegraphics*[width=1\columnwidth]{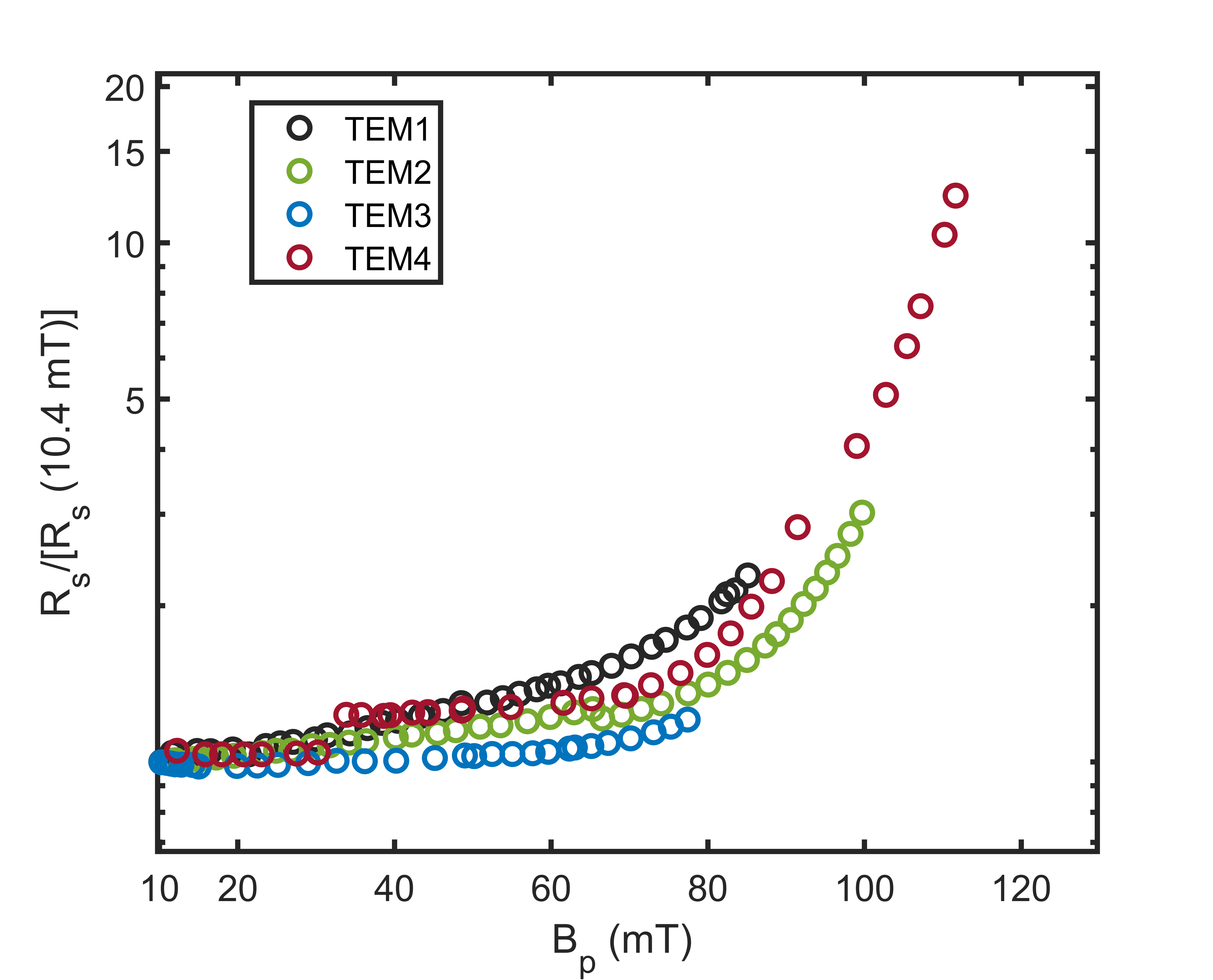}
   \caption{$Q_0(B_p)$ at 2.0 K for first 4 TEM modes. The errors in $B_p$ and $R_s$ are < 5 \% and 10 \%, respectively.}
      \label{Fig3}
\end{figure}

Figure \ref{Fig3} shows the surface resistance normalized to $B_p \sim$ 10 mT as a function of the peak magnetic field for all 4 modes. The cavity quenched at $B_p$ = 112$\pm$5 mT for TEM4 modes with distinct rise in surface resistance starting at $B_p\sim$ 70 mT. For TEM1 and TEM3 modes, the measurements were stopped at $B_p \sim$ 80 mT, whereas the measurement was stopped at $B_p \sim$ 100 mT for TEM2 mode. The RF tests were limited to lower peak field to prevent the additional increase in surface resistance due to residual flux being trapped during cavity quench. 

\subsection{Frequency and $Q$ during warm up}
\begin{figure}[!htb]
   \centering
   \includegraphics*[width=.95\columnwidth]{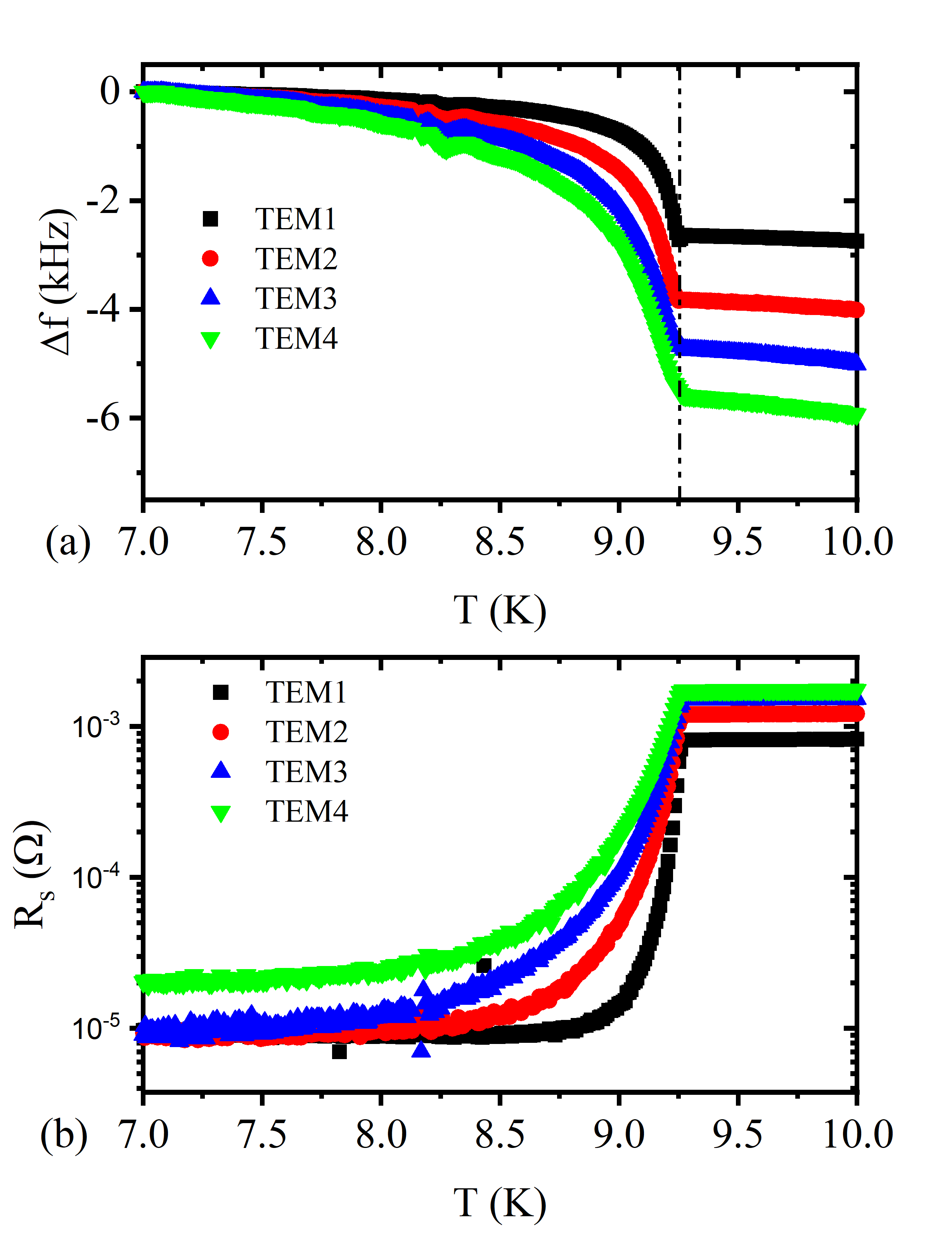}
   \caption{(a) The change in frequency $\Delta f = f-f_{7.0 ~ K}$, and (b) total surface resistance as a function of temperature during warm up. The cavity show the transition to normal conducting state at $T_c \sim$ 9.25 K.}
   \label{fig:warmups}
\end{figure}

The resonant frequency and loaded quality factor of 4 modes were measured during the cavity warm up. The change in frequency is proportional to the penetration depth, whereas the loaded quality factor is converted to the surface resistance $R_s = G/Q_0$ as shown in Fig. \ref{fig:warmups}. The cavity transitioned to superconducting state at $\sim$ 9.25 K. The data in the superconducting state were fitted using the numerical solution of M-B theory \cite{MB} to get the information about the electronic mean free path and found to be 247$\pm$3 nm for TEM1, 248$\pm$4 nm for TEM2, 232$\pm$3 nm for TEM3 and 212$\pm$3 nm, for TEM4 modes. Furthermore, the normal state resistance follows $\sim \sqrt f$ dependence.

\section{Discussion}
\subsection{Frequency dependence analysis}
\begin{figure}[!htp]
   \centering
   \includegraphics*[width=0.98\columnwidth]{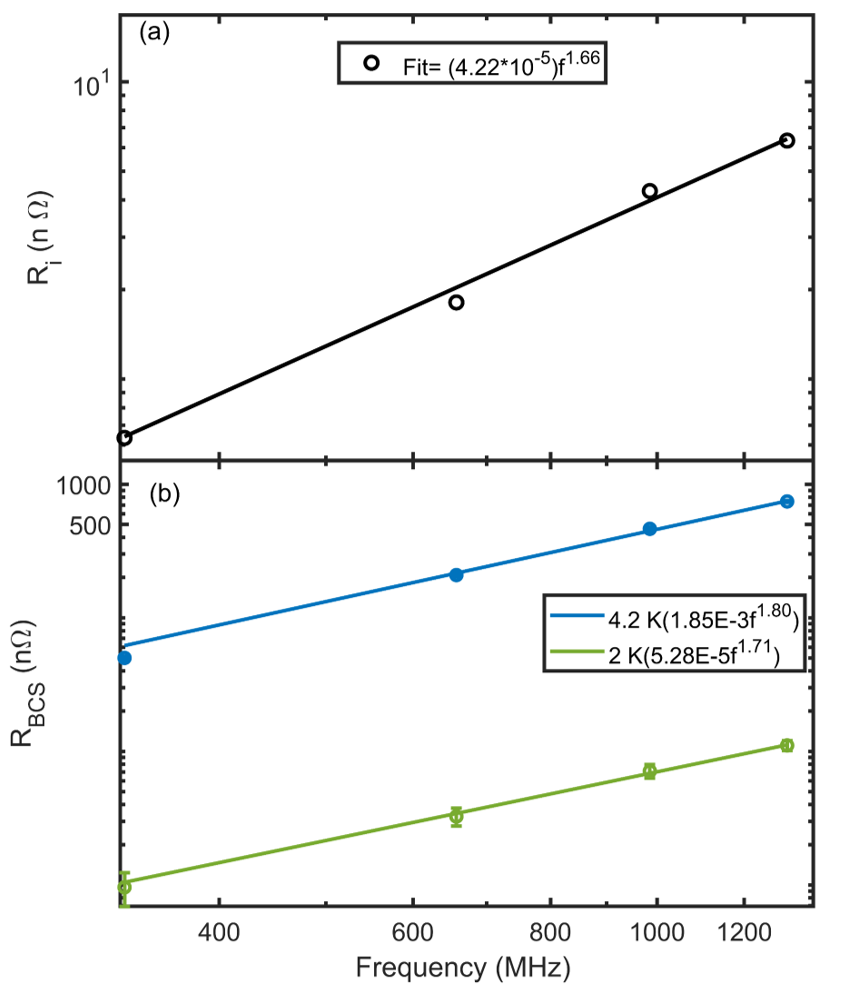}
   \caption{(a) Residual resistance as a function of frequency at $B_p \sim 10 mT $, and (b) $R_{BCS}$ as a function of frequency at 4.2 and 2.0 K.}
   \label{fig:Ri}
\end{figure}

To extract the frequency dependency RF losses, total resistance ($R_s(T)$) measured at $B_p \sim 10 mT$ is fitted using model presented in Ref. \cite{Ciovati}. The residual resistance is independent of the temperature and is dependent on the frequency, trapped magnetic field and purity of the niobium. Figure \ref{fig:Ri} (a) shows the frequency dependence of residual resistance and follows $f^{1.66}$. The frequency dependence of $f^{1.53}$ was previously reported in a similar cavity configuration \cite{parksrf19}. The measurements done on quarter wave and half wave cavity showed an increase in residual resistance with frequency with no clear frequency dependence. Averaging over all the available data, the residual resistance showed $f^{0.7}$ dependence \cite{Kolb}. The large discrepancy in $R_i$ from current study and in Ref. \cite{Kolb} could be due no available data below 2 K showing the asymptotic behaviour in Ref. \cite{Kolb}. Nevertheless, the residual resistance in superconducting state shows strong frequency dependence compared to the anomalous normal conducting loss ($\sim f^{0.5}$). As mentioned, the residual resistance due to trapped flux also has weak frequency dependence ($\sim f^{0.5}$) creating some uncertainty in estimating an exact residual resistance. Further studies are needed to understand the frequency dependence of residual resistance. 

The $R_{BCS}$ resistance is temperature dependent  part of the total resistance ($R_s$) and is plotted in Fig. \ref{fig:Ri} (b) for all four modes. The $R_{BCS}$ resistance was calculated by subtracting $R_i$ from the total surface resistance. The $R_{BCS}$ follows $f^{1.80}$ and $f^{1.71}$ dependence at 4.2 and 2 K, respectively. Exactly same frequency dependence of $f^{1.8}$ at 4.2 K and $f^{1.7}$ at 2.0 K is reported in Ref. \cite{Kolb}. The numerical calculation of $R_{BCS}$ based on the Mattis-Bardeen theory \cite{MB} using the material parameters $\Delta/K_BT_c = 1.85$ and mean free path from the $\Delta\lambda (T)$ fits resulted in $f^{1.79}$ at 4.2 K and $f^{1.77}$ at 2.0 K. The exponent of frequency dependence $R_{BCS}$ agrees well with the calculated values at 4.2 K, however it is $\sim 3$ \% lower at 2.0 K. 
    
\subsection{Calculation of true surface resistance}

\begin{figure}[!htb]
   \centering
   \includegraphics*[width=1\columnwidth]{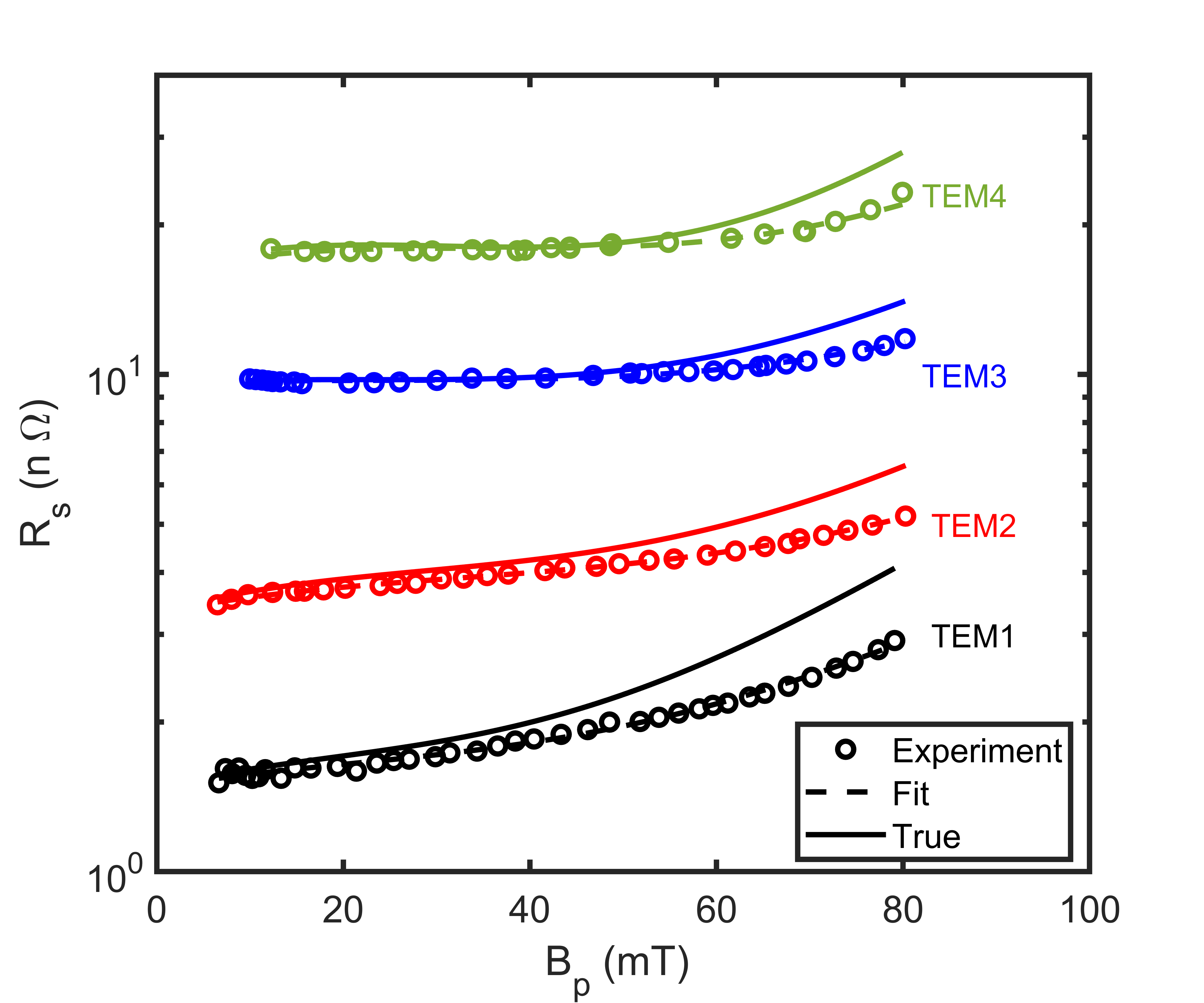}
   \caption{Surface resistance as a function of peak magnetic field of the cavity. The solid lines are the true surface resistance calculated using method developed in Ref. \cite{Delayen}.}
   \label{ActualRs}
\end{figure}

\begin{table*}[t!]
   \centering
   \caption{Summary of fitting parameters. }
   \begin{tabular}{lcccccccc}
       \toprule
\textbf{Modes} & \textbf{$\beta_0$ }  & \textbf{$\beta_1$} & \textbf{$\beta_2$}  & \textbf{$\beta_3$} & \textbf{$R_0$ ($n \Omega$)}  & \textbf{$\alpha_1$} & \textbf{$\alpha_2$}  & \textbf{$\alpha_3$}\\
       \midrule
TEM1    & {1}   & 1.45   & 1.76  & 2.01   & 1.48  & 0.77 &  -1.48  & 2.54 \\ 
TEM2    & {1}   & 1.45   & 1.76  & 2.01   & 3.32  &  0.89 &  -1.64  &  1.74 \\ 
TEM3    & {1}   & 1.45   & 1.76  & 2.01   & 9.51  & 0.25  &  -0.93  &  1.2 \\ 
TEM4     & {1}   & 1.45   & 1.76  & 2.01   &  16.05  & 1.02 &   -2.92 &  2.78  \\ 
       \bottomrule
   \end{tabular}
   \label{table2}
\end{table*}
The surface resistance extracted using $Q_0 = \frac{G}{R_s}$ give an average surface resistance of the cavity, which is a good estimation when the magnetic field is nearly uniform on the cavity surface. For complex cavity geometry like the half-wave coaxial cavity, the non-uniformity of the field distribution makes it harder to estimate the true surface resistance. To get more accurate information about the surface resistance we have fitted the surface resistance $R_s(B_p)$ using the method developed in Ref. \cite{Delayen}. In this method, the experimentally measured surface resistance $R_s (B_p)$ is averaged over the entire surface of the cavity and is expanded as:
\begin{equation}
\overline{R_s}(\frac{B}{B_0}) =\overline{R_0}\sum^{n}_{i=0} \alpha_i(\frac{B}{B_0})^i 
\end{equation}
where,  $\overline{R_0}$ the minimum surface resistance measured and $B_0$  is the maximum peak magnetic field. Then, the real resistance is calculated as:
\begin{equation}
R_{Real}(\frac{B}{B_0}) ={R_0}\sum^{n}_{i=0} \beta(i) \alpha_i(\frac{B}{B_0})^i
\end{equation}
here, $\beta(i)$ is a smooth, continuous, monotonically increasing correction function determined by geometry of the cavity. For the half-wave coaxial cavity, a polynomial of order three (i=3) is a good fit and the fitting parameters are summarized in Table \ref{table2}. As seem from Table \ref{table2}, the experimental data fit well with same sets of $\beta_i$, for given shape. The correction factor of true resistance ($\alpha_i$) differs for different frequency. More analysis will be needed to compare the fitting parameters with respect to frequency, temperature and surface preparations. Figure \ref{ActualRs} shows $R_{Real}$ of all four TEM modes. For $B_p \geq$ 40 mT, the true surface resistance deviate from the experimentally measured average surface resistance for all 4 TEM modes. 

\section{Summary}
We have presented the baseline RF measurements on half-wave coaxial cavity with 4 different TEM modes to understand the frequency dependence of RF properties by minimizing the variability in surface preparation. The  frequency dependence of residual resistance is in consistent with previous results in half-wave coaxial cavity \cite{parksrf15, parksrf17, parksrf19}, but differ from the measurement done using quarter-wave cavity \cite{Kolb}. The BCS surface resistance are consistent with the previous experimental results in both half-wave and quarter-wave cavities \cite{parksrf15, parksrf17, parksrf19, Kolb} as well as the numerical calculation. Future studies will be focused on surface modification with mid-T bake \cite{Lechner}, low temperature baking in UHV \cite{khanal} and nitrogen \cite{dhakal1} as well as the flux trapping sensitivity \cite{dhakal2} to understand the field and frequency dependence of surface resistance. Recently, the cavity received 320 $^\circ$C bake in UHV and RF test will be performed in near future. 

\section{ACKNOWLEDGEMENTS}
We would like to acknowledge Jefferson Lab production staffs for fabrication, processing, assembly,  cryogenic and RF support.

%
\ifboolexpr{bool{jacowbiblatex}}%
	{\printbibliography}%
	{%
	

} 
%
%


\end{document}